\documentclass[
preprint,
amsmath,amssymb, aps, prd, showpacs, nofootinbib ]{revtex4}
\usepackage{graphicx}
\usepackage{dcolumn}
\usepackage{bm}
\usepackage{subfigure}
\usepackage{amsmath, amssymb, graphics}
\usepackage{slashed}
\usepackage{hyperref}
\usepackage[mathlines]{lineno}

\newcommand{\state}[3]{${}^{#1}{#2}_{#3}$}
\newcommand{\stateinequ}[3]{{}^{#1}{#2}_{#3}}

\begin{document}

\title{$O(\alpha_sv^2)$ Corrections to Hadronic and Electromagnetic Decays of \state{1}{S}{0} Heavy
Quarkonium}
\author{Huai-Ke Guo}
\email{huaike.guo@gmail.com}\affiliation{Department of Physics and
State Key Laboratory of Nuclear Physics and Technology, Peking
University, Beijing 100871, China}
\author{Yan-Qing Ma}
\email{yqma.cn@gmail.com}\affiliation{Department of Physics and
State Key Laboratory of Nuclear Physics and Technology, Peking
University, Beijing 100871, China}
\author{Kuang-Ta Chao}
\email{ktchao@th.phy.pku.edu.cn} \affiliation{Department of Physics
and State Key Laboratory of Nuclear
             Physics and Technology, and Center for High Energy Physics,
             \\Peking  University, Beijing 100871, China}
\begin{abstract}
We study $O(\alpha_sv^2)$ corrections to decays of $^1S_0$ heavy
quarkonium into light hadrons and two photons within the framework
of nonrelativistic QCD (NRQCD), and find these $O(\alpha_sv^2)$
corrections to have significant contributions especially for the
decay into light hadrons. With these new results, experimental
measurements of the hadronic width and the $\gamma\gamma$ width of
$\eta_c$ can be described more consistently. By fitting experimental
data, we find the long-distance matrix elements of $\eta_c$ to be
$|\mathcal{R}_{\eta_c}(0)|^2=0.834^{+0.281}_{-0.197}\
\textrm{GeV}^3$ and $\langle \bm{v}^2
\rangle_{\eta_c}=0.232^{+0.121}_{-0.098}$. Moreover, $\eta_c(2S)$ is
also discussed and the $\gamma\gamma$ decay width is predicted to be
$3.34^{+2.06}_{-2.10}\ \textrm{KeV}$.
\end{abstract}

\maketitle

\section{Introduction\label{sec:introduction}}
Heavy quarkonium plays an important role in establishing and
understanding quantum chromodynamics (QCD), the fundamental theory
of strong interactions. Due to the existence of several energy
scales involved with these systems, heavy quarkonium provides an
ideal laboratory for testing the perturbative and nonperturbative
effects of QCD. An effective theory suitable for describing these
systems is nonrelativistic QCD (NRQCD)\cite{nrqed}, which is derived
from QCD by considering the underlying nonrelativistic properties.
According to NRQCD factorization\cite{bbl}, decays of heavy
quarkonium into light hadrons or photons can be organized in a
hierarchy of long-distance matrix elements (LDMEs), which are
classified in terms of $v$, the relative velocity of the heavy
quarks in heavy quarkonium.

Decays of \state{1}{S}{0} heavy quarkonium into light hadrons (LH)
and two photons are among the simplest processes. The short-distance
coefficients for corresponding LDMEs at leading order in $v$ have
been computed previously to next-to-leading order (NLO) in
$\alpha_s$
\cite{Barbieri:1979be,Hagiwara:1980nv,Harris:1957zz,bbl,v4swave,nloquarkonium,huang,1D2,twogamma,lansbergetac,lansbergetab}.
Moreover, that coefficient for $\gamma\gamma$ decay has been
calculated to next-to-next-to-leading order (NNLO) in $\alpha_s$
\cite{nnlo}. However, all coefficients of LDMEs beyond leading order
in $v$ are known at best to leading order in $\alpha_s$
\cite{Keung:1982jb,v4swave,v7,v7gamma}. It is well known that the
calculation at leading order in $\alpha_s$ suffers from large
uncertainties due to strong renormalization scale dependence.
Therefore, to give a more precise description for \state{1}{S}{0}
heavy quarkonium decays beyond leading order in $v$, QCD corrections
to these coefficients are apparently needed.

In this paper, we will study QCD corrections to the coefficients of
order $v^2$ LDME, namely, corrections at order $\alpha_s v^2$ for
\state{1}{S}{0} quarkonium decays. Up to this order of corrections,
there are two unknown LDMEs which should be fixed. Unfortunately,
lattice calculation of these LDMEs \cite{lattice1,lattice2}, though
based on first principles, suffers from large uncertainties. In
Refs.\cite{potentialme,improvedsme,etab}, a new method was
introduced to estimate LDMEs by combining potential models, lattice
calculation, and experimental data. This method will also be used in
this paper to determine the two unknown LDMEs. Then with our
calculated $\alpha_s v^2$ corrections we will be able to get updated
estimates for the decay widths of \state{1}{S}{0} heavy quarkonium
into light hadrons and two photons.

The rest of this paper is organized as follows. We briefly introduce
the theoretical procedures for calculations of heavy quarkonium
decays in Sec.~\ref{sec:theory}. In Sec.~\ref{sec:details}, we
describe kinematics and method of calculation for these processes.
Results in perturbative QCD are summarized in Sec.~\ref{sec:qcd},
while corresponding results in perturbative NRQCD are summarized in
Sec.~\ref{sec:nrqcd}. By using the matching condition, we give the
updated short distance coefficients in Sec.\ref{sec:matching} to
include our new $\alpha_s v^2$ corrections. With these newly
obtained results, we determine the two unknown LDMEs using potential
models and make predictions for relevant decay widths in
Sec.~\ref{sec:phenomenology}. Finally, in Sec.~\ref{sec:summary}, we
present a brief summary.

\section{Decay of Heavy Quarkonium in NRQCD \label{sec:theory}}
The Lagrangian of NRQCD is derived from the QCD Lagrangian by
integrating out the degrees of freedom of order $m_Q$, the mass of
the heavy quark. Local 4-fermion operators are added to accommodate
the inclusive annihilation decay of heavy quarkonium which happens
at scale of order $m_Q$. The Lagrangian of NRQCD is
\begin{equation}
  \mathcal{L}_{\textrm{NRQCD}}=\mathcal{L}_{\textrm{light}}+\mathcal{L}_{\textrm{heavy}}
  +\delta\mathcal{L}.
\end{equation}
Here $\mathcal{L}_{\textrm{heavy}}$ describes nonrelativistic heavy quarks and antiquarks
and is given by
\begin{equation}
  {\mathcal L}_{\textrm {heavy}}
\;=\; \psi^\dagger \, \left( iD_t + \frac{{\bf D}^2}{2 m_Q} \right)\, \psi
\;+\; \chi^\dagger \, \left( iD_t - \frac{{\bf D}^2}{2 m_Q} \right)\, \chi ,
\end{equation}
where $\psi$ is the Pauli spinor field that annihilates a heavy quark,
$\chi$ is the Pauli spinor field  that  creates a heavy antiquark and $D_t$
and ${\bf D}$ are the time and space components of the gauge-covariant derivative
$D^{\mu}$. Terms corresponding to light quarks and gluons are given by
${\mathcal L}_{\textrm {light}}$ and
\begin{equation}
  {\mathcal L}_{\textrm {light}} \;=\; - {1 \over 2} {\textrm {tr}} \, G_{\mu \nu} G^{\mu \nu}
    \;+\; \sum \bar q \; i {\not \! \! D} q ,
\end{equation}
where $G_{\mu \nu}$ is the gluon field strength tensor, $q$ is the Dirac spinor field for light quarks and
the sum is over $n_f$ flavors of light quarks.
Relativistic corrections to the basic effective lagrangian $\mathcal{L}_{\textrm{heavy}}+\mathcal{L}_{\textrm{light}}$
are included in $\delta\mathcal{L}$ and its leading terms are those bilinear in the heavy quark or antiquark field,
\begin{eqnarray}
  \delta{\mathcal L}_{\textrm {bilinear}}
  &=& \frac{c_1}{8m_Q^3}
  \left( \psi^\dagger ({\bf D}^2)^2 \psi \;-\; \chi^\dagger ({\bf D}^2)^2 \chi
  \right)  \nonumber \\
  &+& \frac{c_2}{8m_Q^2}
  \left( \psi^\dagger ({\bf D} \cdot g {\bf E} - g {\bf E} \cdot {\bf D}) \psi
    \;+\; \chi^\dagger ({\bf D} \cdot g {\bf E} - g {\bf E} \cdot {\bf D}) \chi
    \right) \nonumber \\
    &+& \frac{c_3}{8m_Q^2}
    \left( \psi^\dagger (i {\bf D} \times g {\bf E} - g {\bf E} \times i {\bf D})
    \cdot \mbox{\boldmath $\sigma$} \psi
        \;+\; \chi^\dagger (i {\bf D} \times g {\bf E} - g {\bf E} \times i {\bf D})
            \cdot \mbox{\boldmath $\sigma$} \chi \right) \nonumber \\
            &+& \frac{c_4}{2m_Q}
            \left( \psi^\dagger (g {\bf B} \cdot \mbox{\boldmath $\sigma$}) \psi
                \;-\; \chi^\dagger (g {\bf B} \cdot \mbox{\boldmath $\sigma$}) \chi \right),
                  \end{eqnarray}
where $E^i = G^{0i}$ and $B^i = \mbox{$\frac{1}{2}$} \epsilon^{ijk} G^{jk}$
are the electric and magnetic components of the gluon field strength
tensor $G^{\mu \nu}$.

Further corrections include the description of inclusive annihilation decay of heavy
quarkonium and can be achieved by adding local 4-fermion interactions as
\begin{equation}
  \delta{\mathcal L}_{\textrm{4-fermion}}
\;=\; \sum_n {\frac {f_n(\mu_\Lambda)}{ m_Q^{d_n-4}} }{\mathcal O}_n(\mu_\Lambda),
\end{equation}
where $\mu_\Lambda$ is the NRQCD factorization scale, $\mathcal{O}_n(\mu_\Lambda)$ is the local 4-fermion operator,
$d_n$ is the naive scaling dimension of the operator and
$f_n(\mu_\Lambda)$ is the short-distance coefficient which can be
calculated perturbatively.

Thus the decay width of heavy quarkonium can be
given by the following  factorization formula
\begin{equation}
  \Gamma(H)=\sum_n \frac{2\textrm{Im} f_n(\mu_\Lambda) }{m_Q^{d_n-4}}\langle H|\mathcal{O}_n(\mu_\Lambda)|H\rangle,
\end{equation}
where heavy quarkonium state in the Fock space can be written as
\cite{bbl}
\begin{eqnarray}
|H(\stateinequ{2S+1}{L}{J})\rangle&=& O(1)|Q\overline{Q}(\stateinequ{2S+1}{L}{J}^{[1]})\rangle  \nonumber\\
                                  &+& O(v)|Q\overline{Q}(\stateinequ{2S+1}{L\pm 1}{J^{'}}^{[8]})g\rangle \nonumber \\
                                  &+& O(v^2)|Q\overline{Q}(\stateinequ{2S^{'}+1}{L}{J^{'}}^{[8]})g\rangle \\
                                  &+& O(v^2)|Q\overline{Q}(\stateinequ{2S+1}{L}{J}^{[1,8]})gg\rangle \nonumber\\
                                  &+& ... ,\nonumber
\end{eqnarray}
and the relative importance of the 4-fermion operators regarding $v$
can be accessed through the velocity scaling rules
outlined in Ref.\cite{bbl}. We conform to this standard NRQCD power
counting rules throughout this work, although alternative power
counting rules exist\cite{pnrqcd,powercounting1,powercounting2}. A
detailed discussion of the influence of different power counting
rules can be found in Ref.\cite{improvedsme}. For \state{1}{S}{0}
heavy quarkonium decays at order $v^2$, we need only consider the
dominant \state{1}{S}{0} Fock state and two singlet operators with
dimension 6 and 8:
\begin{subequations}
\begin{eqnarray}
\mathcal{O}(\stateinequ{1}{S}{0}^{[1]})&=&\psi^{\dagger}\chi\chi^{\dagger}\psi,\\
\mathcal{P}(\stateinequ{1}{S}{0}^{[1]})&=&{\frac{1}{2}}\left[
\psi^\dagger \chi \chi^\dagger (-\mbox{$\frac{i}{2}$} \tensor{\bf
D})^2 \psi+{\rm h.c.} \right],
\end{eqnarray}
\end{subequations}
for light hadron decay, and
\begin{subequations}
\begin{eqnarray}
  \mathcal{O}_{\textrm{EM}}(\stateinequ{1}{S}{0}^{[1]})&=&\psi^\dagger \chi {| 0 \rangle}{\langle 0 |} \chi^\dagger \psi,\\
  \mathcal{P}_{\textrm{EM}}(\stateinequ{1}{S}{0}^{[1]})&=&{\frac{1}{2}}\left[
\psi^\dagger \chi {| 0 \rangle} {\langle 0 |} \chi^\dagger
(-\mbox{$\frac{i}{2}$} \tensor{\bf D})^2 \psi+{\rm h.c.} \right],
\end{eqnarray}
\end{subequations}
for electromagnetic decay. For a generic color-singlet operator of
the form $\mathcal{O}_n=\psi^\dagger \mathcal{K}_n' \chi
\chi^\dagger \mathcal{K}_n \psi$, applying the Vacuum-Saturation
Approximation \cite{bbl}, we get
\begin{eqnarray}
\langle H |\mathcal{O}_n| H \rangle
&=& \sum_X \langle H | \psi^\dagger \mathcal{K}_n' \chi | X \rangle
    \langle X | \chi^\dagger \mathcal{K}_n \psi | H \rangle
\nonumber \\
&\approx&\langle H | \psi^\dagger \mathcal {K}_n' \chi | 0 \rangle
    \langle 0 | \chi^\dagger \mathcal {K}_n \psi | H \rangle,
\end{eqnarray}
where the omitted terms are of relative order
$v^4$ and are irrelevant of our calculations here. Therefore, in the
following we use the notations
\begin{subequations}
\begin{eqnarray}\label{vacuum}
\langle \mathcal{O}(\stateinequ{1}{S}{0}^{[1]}) \rangle_H:=\langle
H(
\stateinequ{1}{S}{0}^{[1]})|\mathcal{O}(\stateinequ{1}{S}{0}^{[1]})|H(\stateinequ{1}{S}{0}^{[1]})
\rangle &\approx& \langle H(
\stateinequ{1}{S}{0}^{[1]})|\mathcal{O}_{\textrm{EM}}(\stateinequ{1}{S}{0}^{[1]})|H(\stateinequ{1}{S}{0}^{[1]})
\rangle, \\
\langle \mathcal{P}(\stateinequ{1}{S}{0}^{[1]}) \rangle_H:=\langle
H(
\stateinequ{1}{S}{0}^{[1]})|\mathcal{P}(\stateinequ{1}{S}{0}^{[1]})|H(\stateinequ{1}{S}{0}^{[1]})
\rangle &\approx& \langle H(
\stateinequ{1}{S}{0}^{[1]})|\mathcal{P}_{\textrm{EM}}(\stateinequ{1}{S}{0}^{[1]})|H(\stateinequ{1}{S}{0}^{[1]})
\rangle.
\end{eqnarray}
\end{subequations}
Then the  decay width of \state{1}{S}{0} heavy quarkonium at order
$v^2$ is
\begin{subequations}\label{final}
\begin{eqnarray} \label{final1}
  \Gamma(H(\stateinequ{1}{S}{0}^{[1]}) \rightarrow \textrm{LH})&=&
\frac{F(\stateinequ{1}{S}{0}^{[1]})}{m_Q^2}\langle
\mathcal{O}(\stateinequ{1}{S}{0}^{[1]}) \rangle_H
+\frac{G(\stateinequ{1}{S}{0}^{[1]})}{m_Q^4}\langle
\mathcal{P}(\stateinequ{1}{S}{0}^{[1]}) \rangle_H,\\
\label{final2}
 \Gamma(H(\stateinequ{1}{S}{0}^{[1]}) \rightarrow \gamma \gamma)&=&
\frac{F_{\gamma\gamma}(\stateinequ{1}{S}{0}^{[1]})}{m_Q^2}\langle
\mathcal{O}(\stateinequ{1}{S}{0}^{[1]}) \rangle_H
 +\frac{G_{\gamma\gamma}(\stateinequ{1}{S}{0}^{[1]})}{m_Q^4}\langle \mathcal{P}(\stateinequ{1}{S}{0}^{[1]}) \rangle_H,
\end{eqnarray}
\end{subequations}
where the leading order LDME is related to the wave function at
the origin as
\begin{eqnarray} \label{wfodefinition}
\langle \mathcal{O}(\stateinequ{1}{S}{0}^{[1]})
\rangle_H&=&\frac{N_c}{2\pi}|\mathcal{R}_H(0)|^2,
\end{eqnarray}
and a definition of the  ratio of LDMEs is important in this work \cite{improvedsme}
\begin{eqnarray}
\langle \bm{q}^{2r}\rangle_{H}&=& \frac{\langle
0|\chi^\dagger(-\tfrac{i}{2}\tensor{\bm{D}})^{2r}\psi |H\rangle}
{\langle 0|\chi^\dagger\psi|H\rangle},
\end{eqnarray}
where $\bm{q}$ is half the relative momentum of the heavy quark and
anti-quark and it is also convenient to define
\begin{equation}
\label{vsquare}
\langle v^{2r}\rangle_{H}= \langle \bm{q}^{2r}\rangle_{H}/m_Q^{2r}.
\end{equation}
To calculate the short distance coefficients $F$ and $G$ in Eq.
(\ref{final}), we use the matching method \cite{bbl}. Since the
short distance coefficients are insensitive to the long distance
dynamics, we can substitute the bound state with a pair of on shell
quark and antiquark separated by a small relative momentum and
exploit the equivalence of perturbative QCD and perturbative NRQCD
to determine the short-distance coefficients
\begin{equation}
  \label{matching}
\mathcal{A}(Q\overline{Q}\rightarrow
Q\overline{Q})\Big{|}_{\textrm{pert
QCD}}=\sum_{n}\frac{f_{n}(\mu_\Lambda)}{m_Q^{d_{n}-4}}\langle
Q\overline{Q}|\mathcal{O}_{n}
(\mu_\Lambda)|Q\overline{Q}\rangle\Big{|}_{\textrm{pert NRQCD}}\,.
\end{equation}
The left side of this matching equation can be calculated
pertubatively in QCD and the right side can be calculated
perturbatively in NRQCD. Then, we can get the short distance
coefficients $f_{n}(\mu_\Lambda)$ whose imaginary part gives $F$ and
$G$ in Eq.(\ref{final}).

\section{Kinematics and Method of Calculation\label{sec:details}}
We work in the rest frame of the heavy quarkonium and assume the
following notations for the momenta of heavy quark and antiquark
\begin{subequations}
\begin{eqnarray}
p_Q=\frac{1}{2}P+q,\\
p_{\overline{Q}}=\frac{1}{2}P-q,
\end{eqnarray}
\end{subequations}
where
\begin{subequations}
\begin{eqnarray}
P&=&(2E_{\textbf{q}},\mathbf{0}),\\
q&=&(0,\mathbf{q}),
\end{eqnarray}
\end{subequations}
and $E_{\textbf{q}}=\sqrt{m_Q^2+\mathbf{q}^2}$.

In our calculation, we adopt the
covariant spin-projector method \cite{spinprojector0,spinprojector1,spinprojector} to project out the spin-singlet
amplitudes. The projector we use is
\cite{spinprojector}
\begin{equation}
\Pi^{0}=\frac{1}{2\sqrt{2}(E_{\textbf{q}}+m_Q)}(\frac{\slashed{P}}{2}+\slashed{q}+m_Q)
\frac{[(\slashed{P}+2E_{\textbf{q}})\gamma^5(-\slashed{P}+2E_{\textbf{q}})]}{8E_{\textbf{q}}^2}(\frac{\slashed{P}}{2}-\slashed{q}-m_Q).
\end{equation}
To expand the decay width in terms of $\mathbf{q}$, we make the
following rescaling for any momentum $k$,
\begin{equation}
  k \rightarrow k^{'} E_{\textbf{q}}/m_Q,
\end{equation}
which leads all momenta  independent of \textbf{q}, that is, $\partial k_i^{'}\cdot k_j^{'}/\partial \textbf{q}=0$. Thus we can expand the amplitudes in \textbf{q}
before loop integration and phase space integration
and extract the S-wave contribution by making the replacement
\begin{equation}
q_{\mu}q_{\nu} \rightarrow \frac{\mathbf{q}^2}{D-1} [-g_{\mu\nu}+\frac{P_{\mu}^{'}P_{\nu}^{'}}{4m_Q^2}],
\end{equation}
where $P_{\mu}^{'}$ is the rescaled momentum of the heavy quarkonium
which equals $(2m_Q,\mathbf{0})$ in its rest frame. Contributions
coming from potential regions in perturbative QCD and perturbative
NRQCD cancel each other exactly so we neglect these terms to
simplify calculations.

\section{Perturbative QCD Results \label{sec:qcd}}
We use {\tt FeynArts}  \cite{feynarts1,feynarts2} to generate
Feynman diagrams and amplitudes and use self-written {\tt
Mathematica} codes to perform the remained calculations. Ultraviolet
and infrared divergences are regularized with dimensional
regularization and $D=4-2\epsilon$ is assumed. Ultraviolet divergences are
removed by renormalization. We define the renormalized heavy quark mass $m_Q$,
heavy quark field $\psi_Q$ and gluon field $A_{\mu}$ in the on-mass-shell
shceme(OS) and define the QCD coupling constant $g$ in the $\overline{\textrm{MS}}$ scheme, that is,
\begin{eqnarray}
  g^0=Z_g^{\overline{\textrm{MS}}} g,\quad m_Q^0=Z_{m_Q}^{\textrm{OS}} m_Q,\quad
  \psi_Q^0=\sqrt{Z_2^{\textrm{OS}}} \psi_Q,\quad A_{\mu}^0=\sqrt{Z_3^{\textrm{OS}}} A_{\mu},
\end{eqnarray}
where terms with superscript 0 denote bare quantities and $Z_i=1+\delta Z_i$ with $\delta Z_i$ given by
\begin{subequations}
\begin{eqnarray}
  \delta Z_{m_Q}^{\textrm{OS}}&=&-3C_F \frac{\alpha_s}{4\pi}f_{\epsilon}[\frac{1}{\epsilon_{\textrm{UV}}}+\frac{4}{3}+2\ln(2)],\\
  \delta Z_2^{\textrm{OS}}&=&-C_F \frac{\alpha_s}{4\pi}f_{\epsilon}[\frac{1}{\epsilon_{\textrm{UV}}}+\frac{2}{\epsilon_{\textrm{IR}}}+6\ln(2)+4],\\
  \delta Z_3^{\textrm{OS}}&=&\frac{\alpha_s}{4\pi}f_{\epsilon}(\beta_0(n_f)-2C_A)(\frac{1}{\epsilon_{\textrm{UV}}}-\frac{1}{\epsilon_{\textrm{IR}}}),\\
  \delta Z_g^{\overline{\textrm{MS}}}&=&-\frac{\alpha_s}{4\pi}\frac{\beta_0(n_f)}{2}f_{\epsilon}[\frac{1}{\epsilon_{\textrm{UV}}}+\ln(\frac{m_Q^2}{\mu_r^2})+2\ln(2)],
\end{eqnarray}
\end{subequations}
where $f_{\epsilon}=\Gamma(1+\epsilon)[\frac{4\pi
\mu_r^2}{(2m_Q)^2}]^{\epsilon}$, $\beta_0(n_f)=\frac{11}{3}
C_A-\frac{4}{3}T_F n_f$, $\mu_r$ is the renormalization scale and
$n_f$ is the number of light quarks.

\subsection{$\stateinequ{1}{S}{0}^{[1]}\rightarrow \textrm{LH}$}
\begin{figure}
  \begin{center}
    \includegraphics[scale=1]{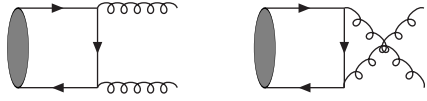}
    \caption{\label{fig1} Feynman diagrams for $\stateinequ{1}{S}{0}^{[1]} \rightarrow gg$ at Born level.}
  \end{center}
\end{figure}

At leading order in $\alpha_s$, there are two diagrams as shown in
Fig.~\ref{fig1}. The corresponding Born level decay width and its
relativistic corrections are
\begin{subequations}
\begin{eqnarray}\label{born}
\Gamma_{\textrm{B}}(\stateinequ{1}{S}{0}^{[1]} \rightarrow gg)&=&
\frac{1}{2!}\frac{1}{2(2m_Q)}\Phi_{(2D)}(\alpha_s 4 \pi)^2
\frac{16}{9m_Q}(1-2\epsilon)(1-\epsilon)\langle\mathcal{O}(\stateinequ{1}{S}{0}^{[1]})\rangle_{\textrm{LO}},\\
\Gamma_{\textrm{B}}^{\textrm{R}}(\stateinequ{1}{S}{0}^{[1]}\rightarrow
gg)&=&-\frac{4}{3}\frac{\mathbf{q}^2}{m_Q^2}\Gamma_{\textrm{B}}(\stateinequ{1}{S}{0}^{[1]}\rightarrow
gg).
\end{eqnarray}
\end{subequations}
where
$\Phi_{(2D)}=\frac{1}{8\pi}\frac{\Gamma(1-\epsilon)}{\Gamma(2-2\epsilon)}(\frac{\pi}{m_Q^2})^{\epsilon}$
is the two-body phase space for $\mathbf{q}=0$ in $D$ dimension. Our
results agree with those in Refs.\cite{1D2,nloquarkonium,v4swave}.
\begin{figure}
\includegraphics[scale=1]{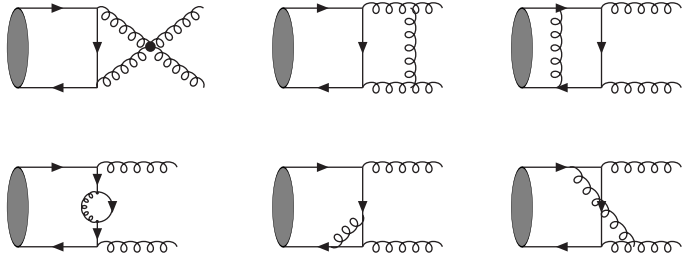}
\caption{\label{fig2}Representative Feynman diagrams for
$\stateinequ{1}{S}{0}^{[1]} \rightarrow gg$ at one-loop level.}
\end{figure}
At next to leading order in $\alpha_s$, there are virtual
corrections and real corrections. Fig.~\ref{fig2} corresponds to
Feynman diagrams of virtual corrections, where only distinct forms
of diagrams are shown. Contribution of these virtual corrections
reads
\begin{eqnarray}\label{virtual}
\begin{split}
  \Gamma_{\textrm{V}}(\stateinequ{1}{S}{0}^{[1]} \rightarrow gg)=&\frac{C_A
  \alpha_s}{\pi}\Gamma_{\textrm{B}}(\stateinequ{1}{S}{0}^{[1]} \rightarrow gg)
f_{\epsilon}
\{ [-\frac{1}{\epsilon^2}+(-\frac{11}{6}+\frac{n_f}{9})\frac{1}{\epsilon} \\
&+\:\frac{1}{36}(-44+19\pi^2+(4n_f-66)\ln(\frac{4m_Q^2}{\mu_r^2}))]\\
&+\:\frac{\mathbf{q}^2}{m_Q^2}[\frac{4}{3}\frac{1}{\epsilon^2}-\frac{4}{27\epsilon}(n_f-31)
+\frac{44}{9}\ln(2) \\
&+\:\frac{1}{324}(-4(11+24\ln(2))n_f+24(33-2n_f)\ln(\frac{m_Q^2}{\mu_r^2})-267\pi^2+874)
] \}.
\end{split}
\end{eqnarray}
While other terms agree with those in Refs.\cite{1D2,nloquarkonium},
the result of relativistic correction here is new. Feynman diagrams
for real corrections are drawn in Fig.~\ref{fig3} and
Fig.~\ref{fig4}, which correspond to final states with three gluons
and $q\overline{q}g$ respectively. Results for these two sets of
real corrections are
\begin{subequations}\label{realLH}
\begin{align}\label{realggg}
\begin{split}
\Gamma(\stateinequ{1}{S}{0}^{[1]} \rightarrow ggg)=&\ \frac{C_A
\alpha_s}{\pi}f_{\epsilon}\Gamma_{\textrm{B}}(\stateinequ{1}{S}{0}^{[1]}
\rightarrow gg)
\{\frac{1}{\epsilon^2}+\frac{11}{6\epsilon}+\frac{181}{18}-\frac{23}{24}\pi^2\\
&+\frac{\mathbf{q}^2}{m_Q^2}[-\frac{4}{3\epsilon^2}-\frac{4}{\epsilon}+\frac{7}{54}(-139+12\pi^2)]\},
\end{split}\\
\label{realqqg} \Gamma(\stateinequ{1}{S}{0}^{[1]} \rightarrow q
\overline{q} g)=&\ n_f\Gamma_{\textrm{B}}(\stateinequ{1}{S}{0}^{[1]}
\rightarrow gg)
\frac{\alpha_s}{\pi}\frac{f_{\epsilon}}{\Gamma(1+\epsilon)\Gamma(1-\epsilon)}T_F(-\frac{2}{3\epsilon}-\frac{16}{9}+
\frac{\mathbf{q}^2}{m_Q^2}(\frac{8}{9\epsilon}+\frac{104}{27})).
\end{align}
\end{subequations}
While other terms agree with those in Refs.\cite{1D2,nloquarkonium},
the results of relativistic corrections here are new. Adding Eqs.
(\ref{virtual}) and (\ref{realLH}), we obtain the  NLO QCD
corrections plus relativistic corrections for the light hadron decay
width of \state{1}{S}{0} heavy quarkonium
  \begin{align}
    \Gamma_{\textrm{QCD}}^{\textrm{NLO}}(\stateinequ{1}{S}{0}^{[1]} \rightarrow \textrm{LH})&=
\frac{C_A
\alpha_s}{\pi}f_{\epsilon}\Gamma_{\textrm{B}}(\stateinequ{1}{S}{0}^{[1]}
\rightarrow gg) \{\frac{1}{216} [
-64n_f+12(2n_f-33)\ln(\frac{4m_Q^2}{\mu_r^2})\nonumber \\
&-93\pi^2+1908 ] +\frac{\mathbf{q}^2}{m_Q^2} [
\frac{16}{27\epsilon} +\frac{1}{324}(24(\ln(\frac{m_Q^2}{\mu_r^2})+2\ln(2))(33-2n_f) \\
&+164n_f+237\pi^2-4964)
]
\nonumber \}.
  \end{align}
Adding these terms together, we get the \state{1}{S}{0} decay width
into light hadrons in perturbative QCD
\begin{equation} \label{qcdlh}
  \Gamma_{\textrm{QCD}}(\stateinequ{1}{S}{0}^{[1]} \rightarrow \textrm{LH})
  =\Gamma_{\textrm{B}}(\stateinequ{1}{S}{0}^{[1]} \rightarrow gg)
  +\Gamma_{\textrm{B}}^{\textrm{R}}(\stateinequ{1}{S}{0}^{[1]}\rightarrow
  gg)
  +\Gamma_{\textrm{QCD}}^{\textrm{NLO}}(\stateinequ{1}{S}{0}^{[1]} \rightarrow \textrm{LH}).
\end{equation}

\begin{figure}
 \begin{center}
    \includegraphics[scale=1]{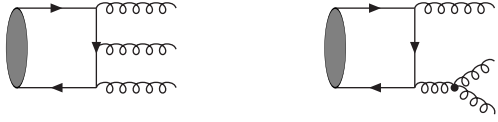}
\caption{\label{fig3}Representative Feynman diagrams for
$\stateinequ{1}{S}{0}^{[1]} \rightarrow ggg$.}
  \end{center}
\end{figure}
\begin{figure}
  \begin{center}
    \includegraphics[scale=1]{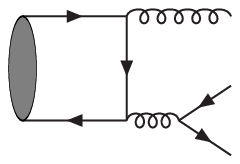}
\caption{\label{fig4}Representative Feynman diagrams for
$\stateinequ{1}{S}{0}^{[1]} \rightarrow q\overline{q}g$.}
  \end{center}
\end{figure}

\subsection{$\stateinequ{1}{S}{0}^{[1]}\rightarrow \gamma \gamma$}
For QCD corrections to the electromagnetic decay, there is no real
correction. Diagrams at Born level and one-loop level are the same
as those in Fig.~\ref{fig1} and Fig~\ref{fig2} except that the final
state gluons are substituted with photons and diagrams containing
three-gluon or four-gluon vertexes are excluded. We then get the
results
\begin{subequations}
\begin{eqnarray}
\label{virtualborn}
\Gamma_{\textrm{B}}(\stateinequ{1}{S}{0}^{[1]} \rightarrow \gamma \gamma)&=&\frac{1}{2!}
\frac{1}{2(2m_Q)}\Phi_{(2D)}(\alpha 4 \pi)^2\frac{8e_Q^4}{m_Q}
\langle\mathcal{O}(\stateinequ{1}{S}{0}^{[1]})\rangle_{\textrm{LO}},\\
\label{virtualre}
\Gamma_{\textrm{B}}^{\textrm{R}}(\stateinequ{1}{S}{0}^{[1]}
\rightarrow \gamma \gamma)&=&-\frac{4}{3}\frac{\mathbf{q}^2}{m_Q^2}
\Gamma_{\textrm{B}}(\stateinequ{1}{S}{0}^{[1]} \rightarrow \gamma \gamma), \\
\label{virtualnew} \Gamma_{\textrm{V}}(\stateinequ{1}{S}{0}^{[1]}
\rightarrow \gamma \gamma)&=&
\Gamma_{\textrm{B}}(\stateinequ{1}{S}{0}^{[1]} \rightarrow \gamma
\gamma)f_{\epsilon}\frac{\alpha_s}{\pi}
[\frac{\pi^2-20}{3}+\frac{\mathbf{q}^2}{m_Q^2}(\frac{16}{9\epsilon}+\frac{196-15\pi^2}{27})],
\end{eqnarray}
\end{subequations}
where $e_Q$ is the electric charge of the heavy quark. Results of
relativistic corrections in Eq. (\ref{virtualnew}) are new and the
other results agree with those previously calculated as summarized
in Ref.\cite{v4swave}. Adding Eqs.(\ref{virtualborn}),
(\ref{virtualre}), and  (\ref{virtualnew}), we get the result for
the $\gamma\gamma$ decay width of \state{1}{S}{0} heavy quarkonium
in perturbative QCD
\begin{equation} \label{qcdgamma}
  \Gamma_{\textrm{QCD}}(\stateinequ{1}{S}{0}^{[1]} \rightarrow \gamma \gamma)=
\Gamma_{\textrm{B}}(\stateinequ{1}{S}{0}^{[1]} \rightarrow \gamma \gamma)
+\Gamma_{\textrm{B}}^{\textrm{R}}(\stateinequ{1}{S}{0}^{[1]} \rightarrow \gamma \gamma)
+\Gamma_{\textrm{V}}(\stateinequ{1}{S}{0}^{[1]} \rightarrow \gamma \gamma).
\end{equation}

\section{Perturbative NRQCD Results\label{sec:nrqcd}}

Order $\alpha_s v^2$ corrections to the leading order LDME
$\langle\mathcal{O}^{0}(\stateinequ{1}{S}{0}^{[1]})\rangle$ in
perturbative NRQCD have been calculated in Ref.\cite{bbl}, where a
cutoff  was introduced to regularize the ultraviolet divergences. We
rewrite it in dimensional regularization,
\begin{equation}
  \langle\mathcal{O}^{0}(\stateinequ{1}{S}{0}^{[1]})\rangle_{\textrm{NLO}}=
  \langle\mathcal{O}^{0}(\stateinequ{1}{S}{0}^{[1]})\rangle_{\textrm{LO}}
[1-\frac{4\alpha_s
C_F}{3\pi}(\frac{\mu_r^2}{\mu_{\Lambda}^2})^{\epsilon}
(\frac{1}{\epsilon_{\textrm{UV}}}-\frac{1}{\epsilon_{\textrm{IR}}})\frac{\mathbf{q}^2}{m_Q^2}].
\end{equation}
We define the renormalized operator
$\mathcal{O}^{\textrm{R}}(\stateinequ{1}{S}{0}^{[1]})$ using the
$\overline{\textrm{MS}}$ scheme
\begin{eqnarray}
  \mathcal{O}^{0}(\stateinequ{1}{S}{0}^{[1]})=Z_O^{\overline{\textrm{MS}}}
  \mathcal{O}^{\textrm{R}}(\stateinequ{1}{S}{0}^{[1]}),
\end{eqnarray}
where
\begin{eqnarray}
  Z_O^{\overline{\textrm{MS}}}=1-\frac{4\alpha_s C_F}{3\pi}(\frac{\mu_r^2}{\mu_{\Lambda}^2})^{\epsilon}(\frac{1}{\epsilon_{\textrm{UV}}}+\ln{4\pi}-\gamma_\textrm{E})
\frac{\mathbf{q}^2}{m_Q^2}.
\end{eqnarray}
Therefore
\begin{eqnarray}
  \langle\mathcal{O}^{\textrm{R}}(\stateinequ{1}{S}{0}^{[1]})\rangle_{\textrm{NLO}}= [1+
\frac{4\alpha_s C_F}{3\pi}(\frac{
\mu_r^2}{\mu_{\Lambda}^2})^{\epsilon}
(\frac{1}{\epsilon}+\ln{4\pi}-\gamma_\textrm{E})
\frac{\mathbf{q}^2}{m_Q^2}]\langle\mathcal{O}(\stateinequ{1}{S}{0}^{[1]})\rangle_{\textrm{LO}}.
\end{eqnarray}
Considering that
  \begin{align}
\langle\mathcal{P}(\stateinequ{1}{S}{0}^{[1]})\rangle_{\textrm{LO}}=
\mathbf{q}^2 \langle\mathcal{O}(\stateinequ{1}{S}{0}^{[1]})\rangle_{\textrm{LO}},
\end{align}
the decay width into light hadrons in perturbative NRQCD becomes
  \begin{align}\label{NRQCD}
  &\Gamma_{\textrm{NRQCD}}(\stateinequ{1}{S}{0}^{[1]}\rightarrow \textrm{LH})= \nonumber\\
&\{F(\stateinequ{1}{S}{0}^{[1]})
+\frac{\mathbf{q}^2}{m_Q^2}[G(\stateinequ{1}{S}{0}^{[1]})+
\frac{4\alpha_s C_F}{3\pi}(\frac{
\mu_r^2}{\mu_{\Lambda}^2})^{\epsilon}(\frac{1}{\epsilon}+\ln{4\pi}-\gamma_\textrm{E})F(\stateinequ{1}{S}{0}^{[1]}
] \}
\frac{\langle\mathcal{O}(\stateinequ{1}{S}{0}^{[1]})\rangle_{\textrm{LO}}}{m_Q^2}.
\end{align}
The electromagnetic decay rate can be obtained by replacing
$F(\stateinequ{1}{S}{0}^{[1]})$ and $G(\stateinequ{1}{S}{0}^{[1]})$
with $F_{\gamma \gamma}(\stateinequ{1}{S}{0}^{[1]})$ and
$G_{\gamma\gamma}(\stateinequ{1}{S}{0}^{[1]})$ respectively.

\section{Matching\label{sec:matching}}
Finally we obtain the short distance coefficients by equating
results from perturbative QCD in Eqs.(\ref{qcdlh}) and
(\ref{qcdgamma}) with that from perturbative NRQCD in Eq.
\eqref{NRQCD}
\begin{subequations}
\begin{align}
F(\stateinequ{1}{S}{0}^{[1]})=&
\frac{4\pi \alpha_s^2}{9}[1+\frac{\alpha_s}{\pi}
\frac{-64n_f+12(2n_f-33)\ln(\frac{4m_Q^2}{\mu_r^2})-93\pi^2+1908}{72}],\\
G(\stateinequ{1}{S}{0}^{[1]})=& \frac{4\pi \alpha_s^2}{9}
\{-\frac{4}{3}+ \frac{\alpha_s}{\pi}\frac{1}{108}
[48\ln(2)(25-2n_f)+164n_f-4964 \nonumber \\
&+24(33-2n_f)\ln(\frac{m_Q^2}{\mu_r^2})+192\ln(\frac{\mu_{\Lambda}^2}{m_Q^2})+237\pi^2]
\},\\
F_{\gamma \gamma}(\stateinequ{1}{S}{0}^{[1]})=& 2\pi \alpha^2 e_Q^4 (1+\frac{\alpha_s}{\pi} \frac{\pi^2-20}{3})
, \\
G_{\gamma \gamma}(\stateinequ{1}{S}{0}^{[1]})=& 2\pi \alpha^2 e_Q^4
\{-\frac{4}{3}+
\frac{\alpha_s}{\pi}\frac{1}{27}[48\ln(\frac{\mu_{\Lambda}^2}{m_Q^2})-96\ln(2)-15\pi^2+196]
\},
\end{align}
\end{subequations}
where QCD corrections for $G(\stateinequ{1}{S}{0}^{[1]})$ and
$G_{\gamma \gamma}(\stateinequ{1}{S}{0}^{[1]})$ are new while the
other results agree with those in Refs.
\cite{1D2,bbl,nloquarkonium,v4swave}. With these short distance
coefficients, we can update the decay widths of \state{1}{S}{0}
heavy quarkonium into light hadrons and two photons.

\section{Phenomenology\label{sec:phenomenology}}
The above obtained  result can be used in \state{1}{S}{0} charmonium
and bottomonium decays. In the following we will focus on the
$\eta_c$ decay width into light hadrons (approximately the total
width) and decay width into two photons. In these decays there are
two unknown LDMEs. In principle, one can fix these LDMEs either
through direct fit with experimental data\cite{he-fan} or
calculation from lattice QCD\cite{lattice1,lattice2}. The order
$v^2$ LDME is ultraviolet divergent and needs to be regularized
\cite{improvedsme}. For lattice calculations this is performed by
imposing a hard cut-off regulator. However, due to slow convergence
of this regularization, the results available from lattice
calculations of order $v^2$ LDME suffer from large
uncertainties\cite{lattice1,lattice2}. On the other hand, we find
that direct fit of the two LDMEs using experimental measurements of
$\gamma\gamma$ width and total width of $\eta_c$ can not give
reliable values due to the approximate linear dependence of the two
theoretical predictions for these two decays.

Therefore we determine the two LDMEs using the potential model
method recently introduced in
Refs.\cite{potentialme,improvedsme,etab}. A widely accepted
potential model, the Cornell potential\cite{charmoniummodel}
\begin{equation}
  V(r)=-\frac{\kappa}{r}+\sigma r,
\end{equation}
is chosen in this work. Since the spin dependent effect is not
included in this potential, the LDMEs calculated this way are
accurate up to corrections of relative order $v^2$. However, as
argued in Ref. \cite{improvedsme}, this error is in fact much less
than the order $v^2$ (about $30 \%$), thus we attach an uncertainty
of $30 \%$ to the central value of the order $v^2$ LDMEs to account
for the error due to this static potential approximation.

In solving the Schr\"odinger equation \cite{schrodinger}, there are
three unknown parameters. $\sigma=0.1682\pm0.0053~\textrm{GeV}^2$ is
taken from the average of lattice calculations\cite{improvedsme} and
the mass parameter is expressed in terms of the $1S$-$2S$ mass
splitting\cite{potentialme,improvedsme}.  Here we take
$m(\psi(2S))-m(J/\psi)=589.188\pm 0.028\ \textrm{MeV}$ \cite{pdg10}.
The last remaining parameter is fixed by equating theoretical
predictions to experimentally measured results. When we use the
decay width formula, we resum a class of relativistic corrections at
leading order in $\alpha_s$ for $\gamma\gamma$ decay as in
Refs.\cite{improvedsme,etab}. For the experimental input, we make
use of this approximation
$\Gamma^{\textrm{LH}}(\eta_c(nS))=\Gamma^{\textrm{total}}(\eta_c(nS))$.
For $\eta_c$, we use $\Gamma^{\gamma\gamma}(\eta_c)$ (or
$\Gamma^{\textrm{LH}}(\eta_c)$) as input to obtain one set LDMEs
which are then utilized to obtain $\Gamma^{\textrm{LH}}(\eta_c)$ (or
$\Gamma^{\gamma\gamma}(\eta_c)$). For $\eta_c(2S)$, we use the total
width as input and make predictions for the $\gamma\gamma$ decay
width. We take $m_c$ to be $1.4\pm0.2 \ \textrm{GeV}$
\cite{improvedsme}, $\alpha=1/137$, $\Lambda_{\textrm{QCD}}=0.39\
\textrm {GeV}$ and vary the renormalization scale $\mu_r$  and NRQCD
factorization scale $\mu_\Lambda$ separately from $1 \ \textrm{GeV}$
to $3\ \textrm{GeV}$ with $2 \ \textrm{GeV}$ as the central value.
The LDMEs are expressed in terms of the wave function at the origin
$|\mathcal{R}(0)|^2$ and $\langle\bm{v}^2\rangle$ as defined in Eqs.
(\ref{wfodefinition}) and (\ref{vsquare}). In each determination of
these LDMEs and the corresponding decay width, we evaluate the
variations caused by the uncertainties of the parameters and
summarize them in the tables. Because the potential does not take
into account of the spin effects, we attach each $v^2$ LDME an
uncertainty $\langle\bm{v}^2 \rangle$ of $30\%$ of the central
value. In each case, various uncertainties are added in quadrature
to give the total uncertainty.

\begin{figure}
      \centering
    \subfigure[$\mu_r$ dependence of $|\mathcal{R}_{\eta_c}^{\gamma\gamma}(0)|^2$]
    {
    \includegraphics[scale=0.8]{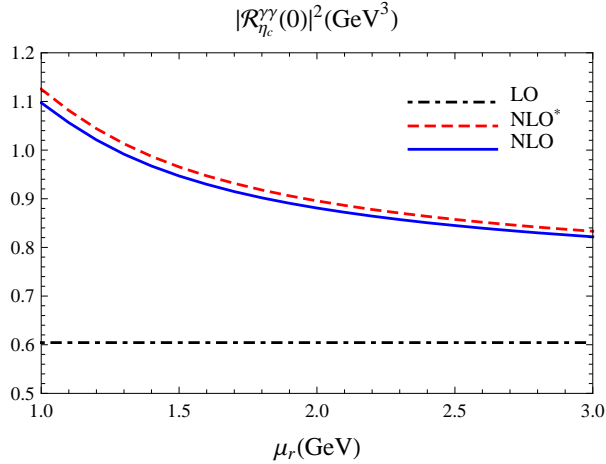}
    }
    \subfigure[$\mu_r$ dependence of $\langle\bm{v}^2 \rangle_{\eta_c}^{\gamma\gamma}$]
    {
    \includegraphics[scale=0.8]{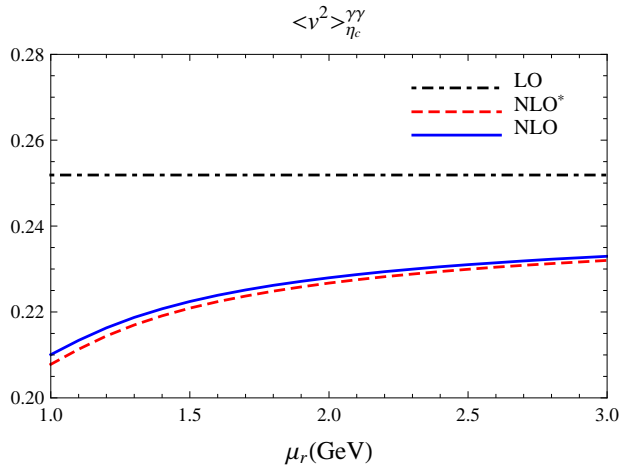}
    }
\caption{\label{fig_me1} $\mu_r$ dependence of LDMEs for $\eta_c$
using the observed $\gamma\gamma$ width as input. LO represents
values without QCD corrections, $\textrm{NLO}^*$ includes QCD
corrections only for terms at leading order in $v$, and NLO takes
into account our new QCD corrections to order $v^2$ terms.}
\end{figure}

For $\eta_c$, with the
$\gamma\gamma$ width $\Gamma^{\gamma\gamma}(\eta_c)=7.2\pm0.7\pm2.0\
\textrm{KeV}$  \cite{pdg10} as input, the determined LDMEs are
\begin{subequations}
\label{etacmegamma}
\begin{eqnarray}
|\mathcal{R}_{\eta_c}^{\gamma\gamma}(0)|^2 &=&
0.881^{+0.382}_{-0.313}\ \textrm{GeV}^3,
\\
\langle\bm{v}^2 \rangle_{\eta_c}^{\gamma\gamma} &=&
    0.228^{+0.126}_{-0.100},
\end{eqnarray}
\end{subequations}
where the superscript $\gamma\gamma$ indicates that we use the
$\gamma\gamma$ decay width as input. Various uncertainties are
summarized in Table~\ref{table1}. The most significant uncertainty
comes from the experimental data. At order $\alpha_s v^2$, another
dependence on NRQCD factorization scale $\mu_\Lambda$ is introduced.
However, as we can see from Table~\ref{table1}, variations of LDMEs
are small when we change $\mu_\Lambda$ from $1\ \textrm{GeV}$ to $3\
\textrm{GeV}$.
 In Fig.~\ref{fig_me1}, we present
$\mu_r$ dependence of three sets of LDMEs in terms of
$|\mathcal{R}_{\eta_c}(0)|^2$ and $\langle\bm{v}^2
\rangle_{\eta_c}$, where the other parameters are fixed to their
central values. Of these three sets of lines, LO represents
calculation without any QCD corrections, $\textrm{NLO}^*$
corresponds to that including QCD corrections but only for terms at
leading order in $v$, and NLO means our new result with order
$\alpha_s v^2$ correction taken into account. The LDMEs
corresponding to $\textrm{NLO}^*$ have been computed earlier in Ref.
\cite{improvedsme} and can be compared here with the values
including the new order $\alpha_s v^2$ corrections. We can see from
this figure that these two lines are close to each other, which
reflects the fact that the effect of the new order $\alpha_s v^2$
correction is not large.
\begin{table}[ht]
\caption{ \label{table1} LDMEs obtained from potential model for
$\eta_c$ and the predicted total width of $\eta_c$, the superscript
$\gamma \gamma$ indicates that the observed width of $\eta_c
\rightarrow \gamma \gamma$ is used as input. The second row gives
central values while the followed rows give variations with respect
to related parameters. }
\begin{ruledtabular}
\begin{tabular}{lccc}
Case&
$|\mathcal{R}_{\eta_c}^{\gamma\gamma}(0)|^2\ (\textrm{GeV}^3)$ &
$\langle\bm{v}^2 \rangle_{\eta_c}^{\gamma\gamma}$ &
$\Gamma^{\textrm{total}}(\eta_c)\ (\textrm{MeV})$
\\
\hline
central&
  0.881 &  0.228 & 31.4\\ $+\Delta \langle \bm{v}^2\rangle_{\eta_c}$&
  0.078 &  0.068 &-3.5 \\ $-\Delta \langle \bm{v}^2\rangle_{\eta_c}$&
 -0.075 & -0.068 & 2.6  \\ $+\Delta m_c$&
  0.187 & -0.065 & 0.3 \\ $-\Delta m_c$&
 -0.167 &  0.102 &-3.4 \\ $+\Delta \sigma$&
  0.022 &  0.020 &-0.9 \\ $-\Delta \sigma$&
 -0.021 & -0.019 & 0.8 \\ $+\Delta \mu_r$&
 -0.059 &  0.005 &-9.2 \\ $-\Delta \mu_r$&
  0.217 & -0.018 & 27.3\\ $+\Delta \mu_\Lambda$&
 -0.036 &  0.003 &-0.6 \\ $-\Delta \mu_\Lambda$&
  0.064 & -0.005 & 1.1 \\ $+\Delta \Gamma_{\eta_c}$&
  0.232 & -0.019 & 10.3\\ $-\Delta \Gamma_{\eta_c}$&
 -0.243 &  0.021 &-9.9
\end{tabular}
\end{ruledtabular}
\end{table}
\begin{figure}
  \centering
  \subfigure[$\mu_r$ dependence of $|\mathcal{R}_{\eta_c}^{\textrm{LH}}(0)|^2$]
    {
    \includegraphics[scale=0.8]{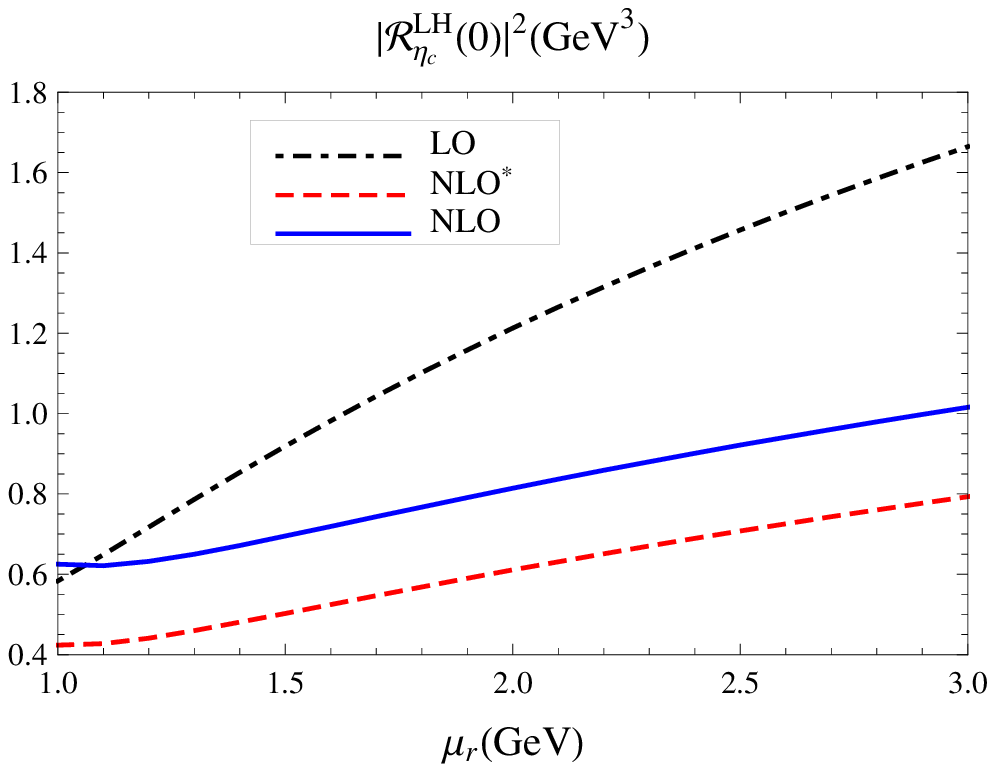}
     \label{fig_me21}
    }
    \subfigure[$\mu_r$ dependence of $\langle\bm{v}^2 \rangle_{\eta_c}^{\textrm{LH}}$]
    {
    \includegraphics[scale=0.8]{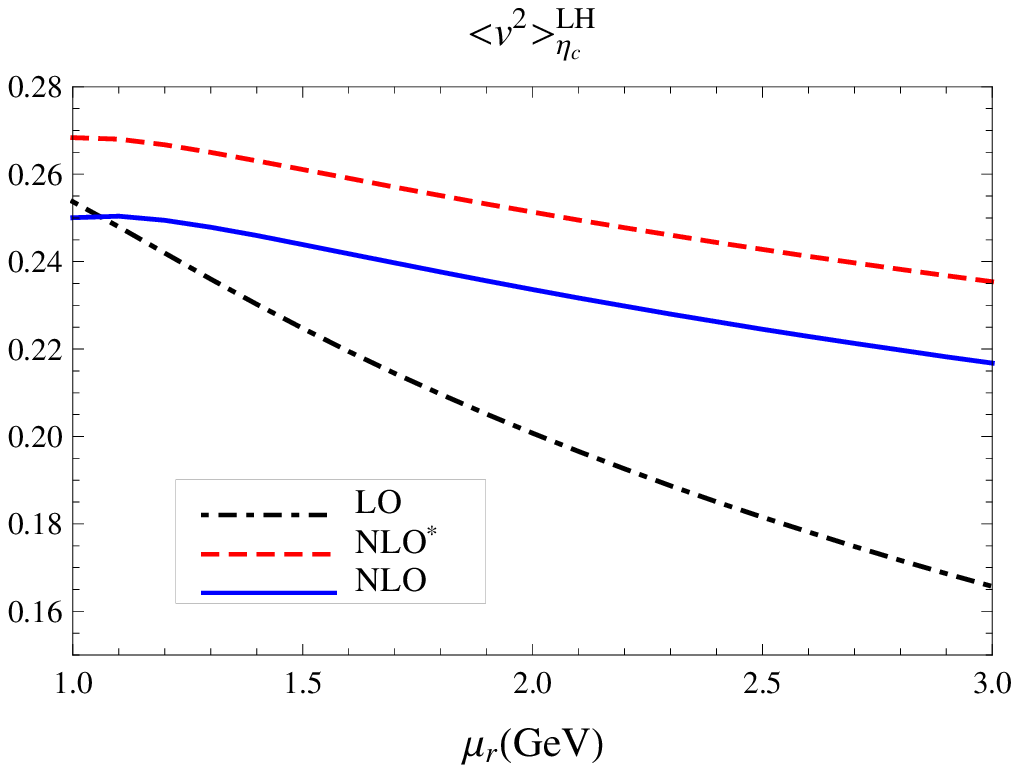}
    \label{fig_me22}
    }
\caption{\label{fig_me2} $\mu_r$ dependence of LDMEs for $\eta_c$
using the observed total width as input. LO represents values
without QCD corrections, $\textrm{NLO}^*$ includes QCD corrections
only for terms at leading order in $v$, and NLO takes into account
our new QCD corrections to order $v^2$ terms.}
  \end{figure}
Utilizing this set of LDMEs as input, we get the total decay width for $\eta_c$
\begin{equation}
  \Gamma^{\textrm{total}}(\eta_c)=31.4^{+29.3}_{-14.4}\  \textrm{MeV}.
  \label{etacwidthgamma}
\end{equation}
This value is in consistency with experimental measurement $28.6\pm
2.2\ \textrm{MeV}$ \cite{pdg10}, although there are large
uncertainties. The details of the uncertainties are summarized in
Table~\ref{table1}. Here, the uncertainty induced by the $\mu_r$
dependence predominates over that from the experimental input of the
$\gamma\gamma$ width of $\eta_c$.
\begin{table}[ht]
\caption{ \label{table2} LDMEs obtained from potential model for
$\eta_c$ and the predicted $\gamma\gamma$ width of $\eta_c$, where
the superscript LH indicates that the observed total width of
$\eta_c$ is used as input. The second row gives the central values
while the followed rows give variations with respect to related
parameters. }
\begin{ruledtabular}
\begin{tabular}{lccc}
Case&
$|\mathcal{R}_{\eta_c}^{\textrm{LH}}(0)|^2\ (\textrm{GeV}^3)$ &
$\langle\bm{v}^2 \rangle_{\eta_c}^{\textrm{LH}}$  &
$\Gamma(\eta_c\rightarrow \gamma\gamma)\ (\textrm{KeV})$
\\
\hline
central&
  0.814 &  0.234 & 6.61\\ $+\Delta \langle \bm{v}^2\rangle_{\eta_c}$&
  0.194 &  0.070 & 0.90\\ $-\Delta \langle \bm{v}^2\rangle_{\eta_c}$&
 -0.131 & -0.070 &-0.54 \\ $+\Delta m_c$&
  0.163 & -0.065 &-0.07\\ $-\Delta m_c$&
 -0.090 &  0.095 & 0.71\\ $+\Delta \sigma$&
  0.042 &  0.018 & 0.19 \\ $-\Delta \sigma$&
 -0.038 & -0.018 &-0.16 \\ $+\Delta \mu_r$&
  0.201 & -0.017 & 2.47\\ $-\Delta \mu_r$&
 -0.189 &  0.016 &-2.73\\ $+\Delta \mu_\Lambda$&
 -0.020 &  0.002 & 0.13\\ $-\Delta \mu_\Lambda$&
  0.036 & -0.003 &-0.21\\ $+\Delta \Gamma_{\eta_c}$&
  0.052 & -0.004 & 0.46\\ $-\Delta \Gamma_{\eta_c}$&
 -0.053 &  0.005 &-0.47
\end{tabular}
\end{ruledtabular}
\end{table}

If, on the other hand, we use the total width of $\eta_c$,
$\Gamma^{\textrm{total}}(\eta_c)=28.6\pm2.2\ \textrm{MeV}$
\cite{pdg10} as input, then we get another set of LDMEs
\begin{subequations}
\label{etacmegg}
\begin{eqnarray}
  |\mathcal{R}_{\eta_c}^{\textrm{LH}}(0)|^2 &=&
     0.814^{+0.332}_{-0.256}\ \textrm{GeV}^3,
\\
\langle\bm{v}^2 \rangle_{\eta_c}^{\textrm{LH}} &=&
     0.234^{+0.121}_{-0.099},
\end{eqnarray}
\end{subequations}
where the superscript LH indicates that we use the total width of
$\eta_c$ as input. Variations with respect to the parameters are
summarized in Table~\ref{table2}. The experimental uncertainty in
this case is small and the main uncertainty comes from the
relatively strong $\mu_r$ dependence of the theoretical prediction.
The $\mu_r$ dependence of the two LDMEs is shown in
Fig.~\ref{fig_me2}. As in previous case, we display another two sets
of LDMEs, where only QCD corrections at leading order in $v$ are
taken into account or no QCD correction is considered. The two sets
with QCD corrections show great improvement of $\mu_r$ dependence
with respect to the one without QCD corrections, and they are almost
parallel to each other. The only difference between the two sets of
values is that the $\alpha_s v^2$  correction enhances
$|\mathcal{R}_{\eta_c}(0)|^2$ by about $30\%$. This enhanced LDME
coincides with previously obtained value using
$\Gamma^{\gamma\gamma}(\eta_c)$ as input. The value of
$\langle\bm{v}^2 \rangle_{\eta_c}$ is relatively stable. With this
set of LDMEs, we obtain the $\gamma\gamma$ decay width
\begin{equation}
  \Gamma^{\gamma\gamma}(\eta_c)=6.61^{+2.77}_{-2.83}\ \textrm{KeV},
  \label{etacwidthgamma}
\end{equation}
which is also consistent with the experimental measurement
$7.2\pm0.7\pm2.0\ \textrm{KeV}$ \cite{pdg10}.

Since now we have two sets of values for the two LDMEs for $\eta_c$
in Eqs.(\ref{etacmegamma}) and (\ref{etacmegg}), we can combine
these values to get a better estimation. The
uncertaintes in Table~\ref{table1} and Table~\ref{table2} are correlated,
and we use the method in Ref.~\cite{improvedsme} to treat these correlations.
First we construct a two-by-two covariance matrix for
$\langle \mathcal{O}(\stateinequ{1}{S}{0}^{[1]}) \rangle_{\eta_c}^{\gamma\gamma}$ and
$\langle \mathcal{O}(\stateinequ{1}{S}{0}^{[1]}) \rangle_{\eta_c}^{\textrm{LH}}$.
It describes correlations between the variations in the two tables
and is defined as $C_{jk}=\sum_{i} \Delta_{ji}\Delta_{ki}$ with $\Delta_{ji}=\frac{1}{2}(O_{ji}^+-O_{ji}^-)$.
The indexes $j,k$ refer to these two leading order LDMEs and
$i$ runs through every item in Table~\ref{table1} and Table~\ref{table2}. $O_{ji}^+$ and $O_{ji}^-$ correspond to the
plus and minus variations of the LDMEs.
For the $i$-th item in Table~\ref{table1} or Table~\ref{table2}, we define the $\chi^2_i$ as
\begin{eqnarray}
  \chi^2_i=\sum_{j,k}(O_{ji}-\overline{O}_{i})(C^{-1})_{jk}(O_{ki}-\overline{O}_{i})
\end{eqnarray}
and minimize it to get the average value $\overline{O}_{i}$ for
$\langle \mathcal{O}(\stateinequ{1}{S}{0}^{[1]})_{\eta_c}$. Once we obtain
the values of $\langle \mathcal{O}(\stateinequ{1}{S}{0}^{[1]})_{\eta_c}$, we
use the potential model to get the values of $\langle \mathcal{P}(\stateinequ{1}{S}{0}^{[1]})_{\eta_c}$.
We perform this calculation for each of the
items in Table~\ref{table1} and  Table~\ref{table2} , treat the renormalization
scale $\mu_r$ and NRQCD factorization scale $\mu_{\Lambda}$ simply as the
same quantities in the two tables and express the results in terms of $|\mathcal{R}_{\eta_c}(0)|$ and $\langle \bm{v}^2 \rangle_{\eta_c}$.
The results are
\begin{subequations}
  \begin{eqnarray}
  |\mathcal{R}_{\eta_c}(0)|^2&=&0.834^{+0.281}_{-0.197}\ \textrm{GeV}^3, \\
  \langle \bm{v}^2 \rangle_{\eta_c}&=&0.232^{+0.121}_{-0.098},
\end{eqnarray}
\end{subequations}
with the details of the uncertainties given in Table~\ref{table12}.
We note that the uncertainties here for $|\mathcal{R}_{\eta_c}(0)|^2$ are smaller than those in
Eqs.(\ref{etacmegamma}) and (\ref{etacmegg}) .
\begin{table}[ht]
\caption{ \label{table12} The averages of the LDMEs for $\eta_c$. The second row gives the central values and subsequent rows
give variations with respect to various uncertainties.}
\begin{ruledtabular}
\begin{tabular}{lccc}
Case&
$|\mathcal{R}_{\eta_c}(0)|^2\ (\textrm{GeV}^3)$ &
$\langle\bm{v}^2 \rangle_{\eta_c}$
\\
\hline
central&
  0.834 &  0.232 \\ $+\Delta \langle \bm{v}^2\rangle_{\eta_c}$&
  0.159 &  0.070 \\ $-\Delta \langle \bm{v}^2\rangle_{\eta_c}$&
 -0.115 & -0.070  \\ $+\Delta m_c$&
  0.170 & -0.065  \\ $-\Delta m_c$&
 -0.113 &  0.097  \\ $+\Delta \sigma$&
  0.036 &  0.019  \\ $-\Delta \sigma$&
 -0.033 & -0.018  \\ $+\Delta \mu_r$&
  0.124 & -0.010  \\ $-\Delta \mu_r$&
 -0.069 &  0.006 \\ $+\Delta \mu_\Lambda$&
 -0.025 &  0.002 \\ $-\Delta \mu_\Lambda$&
  0.044 & -0.004 \\ $+\Delta \Gamma_{\eta_c}^{\gamma\gamma}$&
  0.069 & -0.006 \\ $-\Delta \Gamma_{\eta_c}^{\gamma\gamma}$&
 -0.072 &  0.006 \\ $+\Delta \Gamma_{\eta_c}^{total}$&
  0.037 & -0.003 \\ $-\Delta \Gamma_{\eta_c}^{total}$&
 -0.037 &  0.003
\end{tabular}
\end{ruledtabular}
\end{table}

\begin{figure}
  \centering
  \subfigure[$\mu_r$ dependence of $|\mathcal{R}_{\eta_c(2S)}^{\textrm{LH}}(0)|^2$]
    {
    \includegraphics[scale=0.8]{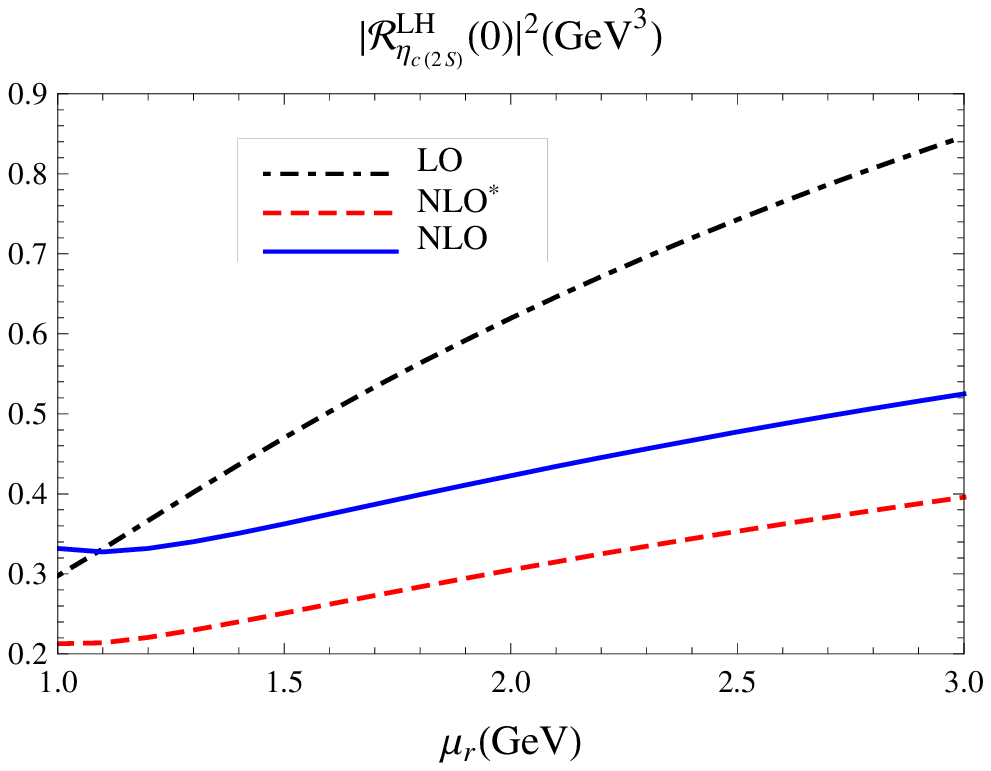}
    \label{fig_me31}
    }
    \subfigure[$\mu_r$ dependence of $\langle\bm{v}^2 \rangle_{\eta_c(2S)}^{\textrm{LH}}$]
    {
    \includegraphics[scale=0.8]{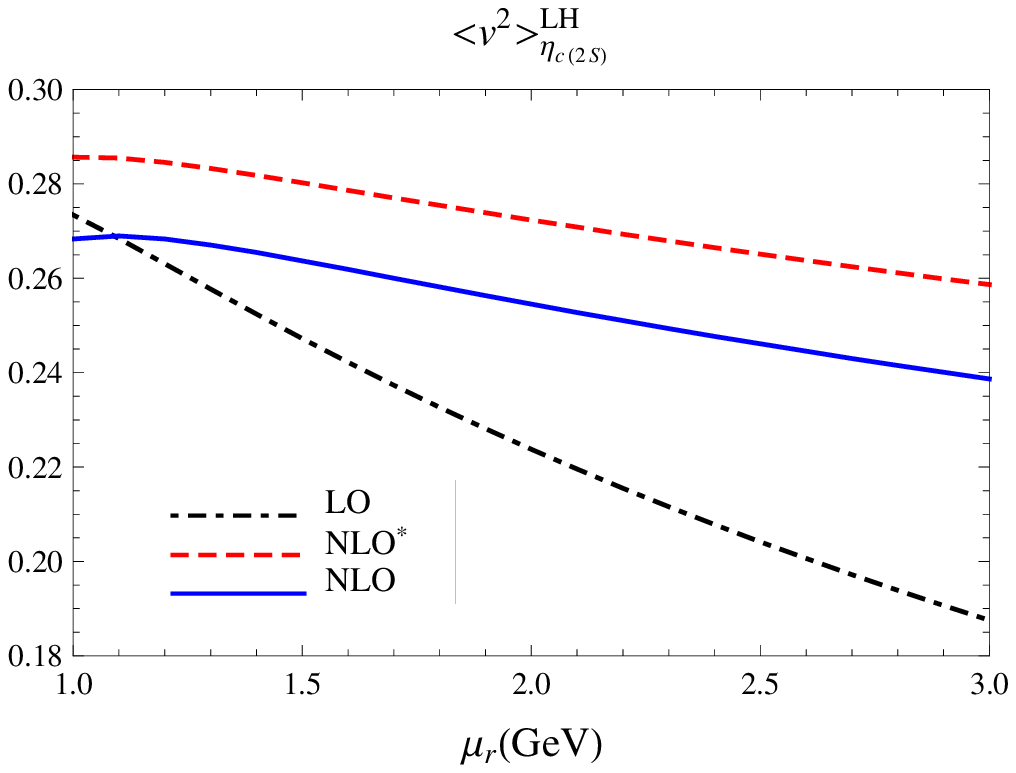}
    \label{fig_me32}
    }
\caption{\label{fig_me3} $\mu_r$ dependence of LDMEs for
$\eta_c(2S)$ using the total width as input. LO represents values
without QCD corrections, $\textrm{NLO}^*$ includes QCD corrections
only for terms at leading order in $v$, and NLO takes into account
our new QCD corrections to order $v^2$ terms.}
  \end{figure}

For $\eta_c(2S)$, we use the observed total width $14\pm 7 \
\textrm{Mev}$\cite{pdg10} as input and get the LDMEs for
$\eta_c(2S)$
\begin{subequations}
\label{etac2smegg}
\begin{eqnarray}
  |\mathcal{R}_{\eta_c(2S)}^{\textrm{LH}}(0)|^2 &=&
     0.423^{+0.245}_{-0.230}\ \textrm{GeV}^3,
\\
\langle\bm{v}^2 \rangle_{\eta_c(2S)}^{\textrm{LH}} &=&
     0.255^{+0.130}_{-0.109}.
\end{eqnarray}
\end{subequations}
Some potential model calculations of the squared wave function at
the origin in Ref.\cite{wfo} give 0.418 for the logarithmic
potential\cite{logpotential}, 0.529 for the QCD-motivated B-T
model\cite{btpotential}, and 0.559 for the power-law
potential\cite{powerpotential}. Our result is consistent with their
values. Table~\ref{table3} gives details with various uncertainties.
In the $\eta_c(2S)$ case, both the $\mu_r$ dependence of theoretical
result and the experimental input of the total width have large
uncertainties and therefore the LDMEs are subject to relatively
large uncertainties. In Fig.~\ref{fig_me3}, we present $\mu_r$
dependence of this set of LDMEs. The shape of the lines is similar
to Fig.~\ref{fig_me2} except the value for
$|\mathcal{R}_{\eta_c(2S)}(0)|^2$ here is smaller by about a factor of
2.
\begin{table}[ht]
\caption{ \label{table3} LDMEs obtained from potential model for
$\eta_c(2S)$ and the predicted $\gamma\gamma$ width of $\eta_c(2S)$,
where the superscript LH indicates that the total width of
$\eta_c(2S)$ is used as input. The second row gives the central
values while the followed rows give variations with respect to
related parameters. }
\begin{ruledtabular}
\begin{tabular}{lccc}
Case&
$|\mathcal{R}_{\eta_c}^{\textrm{LH}}(0)|^2\ (\textrm{GeV}^3)$ &
$\langle\bm{v}^2 \rangle_{\eta_c}^{\textrm{LH}}$ &
$\Gamma(\eta_c(2S)\rightarrow \gamma \gamma)\ (\textrm{KeV})$
\\
\hline
central&
 0.423 & 0.255 & 3.34 \\ $+\Delta \langle \bm{v}^2\rangle_{\eta_c(2S)}$&
 0.121 & 0.076 & 0.58\\ $-\Delta \langle \bm{v}^2\rangle_{\eta_c(2S)}$&
-0.077 &-0.076 &-0.33\\ $+\Delta m_c$&
 0.077 &-0.069 &-0.08\\ $-\Delta m_c$&
-0.035 & 0.099 & 0.47\\ $+\Delta \sigma$&
 0.024 & 0.019 & 0.11\\ $-\Delta \sigma$&
-0.022 &-0.019 &-0.10\\ $+\Delta \mu_r$&
 0.102 &-0.016 & 1.22\\ $-\Delta \mu_r$&
-0.091 & 0.014 &-1.33\\ $+\Delta \mu_\Lambda$&
-0.012 & 0.002 & 0.07\\ $-\Delta \mu_\Lambda$&
 0.022 &-0.003 &-0.11\\ $+\Delta \Gamma_{\eta_c(2S)}$&
 0.167 &-0.026 & 1.47\\ $-\Delta \Gamma_{\eta_c(2S)}$&
-0.192 & 0.029 &-1.58
\end{tabular}
\end{ruledtabular}
\end{table}
Exploiting this set of LDMEs, we can make predictions for the
$\gamma\gamma$ decay width of $\eta_c(2S)$,
\begin{equation}
  \Gamma^{\gamma\gamma}(\eta_c(2S))=3.34^{+2.06}_{-2.10}\ \textrm{KeV} .
  \label{etacwidthgamma}
\end{equation}
This prediction is consistent with the experimental observation that
the branching fraction of $\gamma\gamma$ decay is smaller than
$5\times 10^{-4}$ \cite{pdg10}. Another experiment measured
$\Gamma_{\gamma \gamma}(\eta_c(2S))B(\eta_c(2S)\rightarrow K
\overline{K}\pi) = (0.18\pm 0.05\pm 0.02)\Gamma_{\gamma
\gamma}(\eta_c(1S)) B(\eta_c(1S)\rightarrow K \overline{K}\pi)$
\cite{etac2sgamma} and assumed that the branching fractions of
$\eta_c$ and $\eta_c(2S)$ decays into $K_S K \pi$ were equal, and
made use of $\Gamma(\eta_c\rightarrow \gamma\gamma)=7.4\pm0.4\pm2.3\
\textrm{KeV}$, and then derived $\Gamma(\eta_c(2S))=1.3\pm 0.6 \
\textrm{KeV}$\cite{pdg10,etac2sgamma}. Our result is not in
contradiction with their measurement within errors.

\section{Summary\label{sec:summary}}

Within the framework of NRQCD, we calculate order  $\alpha_s v^2$
corrections to decays of \state{1}{S}{0} heavy quarkonium  into
light hadrons and two photons. In both processes, infrared
divergences are found to be canceled through the matching of
perturbative QCD and perturbative NRQCD results. There are two
unknown NRQCD LDMEs, which are determined using potential model
method\cite{potentialme,improvedsme,etab} either with the observed
total width or two photon width as input. When using
$\Gamma^{\gamma\gamma}(\eta_c)$ as input, we get
$|\mathcal{R}_{\eta_c}^{\gamma\gamma}(0)|^2=0.881^{+0.382}_{-0.313}\
\textrm{GeV}^3$ and $\langle\bm{v}^2
\rangle_{\eta_c}^{\gamma\gamma}= 0.228^{+0.126}_{-0.100}$, from
which we predict
$\Gamma^{\textrm{total}}(\eta_c)=31.4^{+29.3}_{-14.4}\
\textrm{MeV}$. Alternatively, when using
$\Gamma^{\textrm{total}}(\eta_c)$ as input, we get
$|\mathcal{R}_{\eta_c}^{\textrm{LH}}(0)|^2=0.814^{+0.332}_{-0.256}\
\textrm{GeV}^3$ and $\langle\bm{v}^2 \rangle_{\eta_c}^{\textrm{LH}}=
0.234^{+0.121}_{-0.099}$, and we predict the $\gamma\gamma$ width of
$\eta_c$ to be $\Gamma^{\gamma\gamma}(\eta_c)=6.61^{+2.77}_{-2.83}\
\textrm{KeV}$. All these predictions agree well with experimental
data. We then combine these two kinds of determination of LDMEs and
get the average values
$|\mathcal{R}_{\eta_c}(0)|^2=0.834^{+0.281}_{-0.197}\
\textrm{GeV}^3$ and $\langle\bm{v}^2 \rangle_{\eta_c}=
0.232^{+0.121}_{-0.098}$. For $\eta_c(2S)$, we use the observed
total width as input and find
$|\mathcal{R}_{\eta_c(2S)}^{\textrm{LH}}(0)|^2=0.423^{+0.245}_{-0.230}\
\textrm{GeV}^3$ and $\langle\bm{v}^2
\rangle_{\eta_c(2S)}^{\textrm{LH}}= 0.255^{+0.130}_{-0.109}$. With
this set of LDMEs, we predict the $\gamma \gamma$ width of
$\eta_c(2S)$ to be $3.34^{+2.06}_{-2.10}\ \textrm{KeV}$, which is
not in contradiction with data within uncertainties. Consequently,
the order $\alpha_s v^2$ corrections (especially the one to the
decay into light hadrons) are found to have significant effects on
improving the consistency between theoretical predictions and
experimental measurements.

$Note\ added$. When we finished the calculations and are preparing
this paper, a related work appears\cite{jiayu} that also gives
$\alpha_s v^2$ corrections to the $\gamma \gamma$ decay width. We
find our results for this channel agree with theirs, while we have
also calculated the light hadron decay width.

\section*{Acknowledgement}
We thank G. T. Bodwin, Y. Fan, and C. Meng for helpful discussions.
We also thank Franz F. Sch\"oberl for providing us with
schroedinger.nb to solve the Schr\"odinger equation. This work was
supported by the National Natural Science Foundation of China
(No.11021092, No.11075002) and the Ministry of Science and
Technology of China (No.2009CB825200).
\providecommand{\href}[2]{#2}\begingroup\raggedright\endgroup

\end{document}